# Nitrogen Contamination in Elastic Neutron Scattering

Songxue Chi[1,2], Jeffrey W. Lynn[1], Ying Chen[1,2], William Ratcliff II[1], Benjamin G. Ueland[1], Nicholas P. Butch[3], Shanta R. Saha[3], Kevin Kirshenbaum[3] and Johnpierre Paglione[3]

[1]NIST Center for Neutron Research, National Institute of Standards and Technology, Gaithersburg, Maryland 20899, USA
[2]Department of Materials Science and Engineering, University of Maryland, College Park, Maryland 20742, USA
[3]CNAM and Department of Physics, University of Maryland, College Park, Maryland 20742, USA

**Abstract**

Nitrogen gas accidentally sealed in a sample container produces various spurious effects in elastic neutron scattering measurements. These effects are systematically investigated and the details of the spurious scattering are presented.

## 1. Introduction

The most commonly observed effect of air scattering during a neutron scattering measurement is a contribution to the background, especially in inelastic measurements which require longer counting times. Another source of 'background', by which we really mean unwanted and/or unexpected cross sections observed in a measurement, is when exchange gas is used in sample environment equipment. Typically helium is employed as the exchange gas because it remains mobile for thermal conduction purposes at all temperatures (if sufficient helium is present) and has a small neutron cross section. Nevertheless, the scattering from liquid helium can be observed at low temperatures, below ≈4 K[1,2]. However, contamination by air, in particular diatomic nitrogen which scatters much more strongly than helium, can occur if the source of He itself is contaminated, the system used to introduce the exchange gas contains some air, or the sample environment system leaks air into the sample area. In this situation one might see changes in the background scattering at much higher temperatures, somewhere below 77 K depending on the partial pressure of $N_2$ [1]. Moreover, as a classical diatomic van der Walls crystal, solid nitrogen exhibits structural transitions [3] that can also cause sharp changes in the background as well as peaks in the scattering that can be misidentified. Here we report some of the effects of having $N_2$ in the sample environment system during elastic neutron scattering measurements. The spurious changes observed include possible sudden changes in the Bragg intensities depending on how the sample is mounted, and the development of new Bragg peaks, in addition to the usual changes in 'background' scattering that is usually more obvious in inelastic neutron scattering. The anomalous changes in the observed Bragg scattering are found to occur on warming but not on cooling, and furthermore depend on rate of change of the temperature. The overall behaviour can make it difficult to readily identify this scattering as spurious.

## 2. Experimental methods

The sample used for this study is a 21 mg $SrFe_{1.92}Ni_{0.08}As_2$ single crystal, and its tetragonal lattice parameters are $a$ = 5.5746 Å, $b$ = 5.513 Å, and $c$ = 12.286 Å at T=60 K. The sample was sealed in an aluminium can in a glove box with He as the intended exchange gas. This common practice of sealing was found to be prone to $N_2$ contamination because even after purging, the glove box still contained enough $N_2$ to cause spurious effects in elastic scattering. The $N_2$ containing sample can was then mounted in a bottom-loading closed cycle refrigerator (CCR). Neutron diffraction measurements were carried out on the BT-7 and BT-9 triple axis spectrometers at the NIST Center for Neutron Research. A typical instrumental configuration was employed, using a neutron energy of 14.7 meV provided by pyrolytic graphite crystals as monochromator, analyzer

(when employed) and filter. For the measurements presented, Söller collimations of 60′-50′-S-50′ (with no analyzer) were used on BT-7 and 40′-47′-S-40′-100′ were used on BT-9.

## 3. Results and discussion

Solid nitrogen is known to have at least five phases up to 20 GPa: α, β, γ, δ, and ε-phases [3]. The γ-, δ- and ε-phases appear when pressure is applied and therefore are not of interest here. The crystal structure is cubic (Pa3) for α-$N_2$ and hexagonal ($P6_3/mmc$) for β-$N_2$, and at equilibrium vapour pressure the α↔β phase transition occurs at $T_{\alpha \leftrightarrow \beta}$ =36 K [4]. This is a first order transition and is accompanied by a 1% volume change. The normal melting and boiling points of $N_2$ at 1 atm are about 63 K and 77 K respectively.

With inadvertent amount of $N_2$ sealed in the sample can, we observed a few concentric powder rings from solid $N_2$. One such powder ring of scattering at 50K is shown in figure 1(a), together with the magnetic Bragg peak (1,0,3) from the $SrFe_{1.92}Ni_{0.08}As_2$ sample. The ring occurs at $|\mathbf{Q}|$=2.12 Å$^{-1}$ and can be identified as the (1,0,1) peak of hexagonal β-$N_2$ which corresponds to $d$=2.964 Å. The intensity of the scattering on the ring position is shown in figure 2(a) as a function of temperature. We see that upon warming the scattering starts to develop around 30 K when the temperature is changed by 0.2 K for a counting time of 20 s, while it starts to develop around 40 K when warming at the faster rate of 0.5 K every 20 s. The powder rings of scattering abruptly disappear at ≈60 K. Interestingly the powder ring appears on warming, but not on cooling as seen in figure 1(b) and figure 2(a). We expect the reason for this is because when cooling, the container itself is at a lower temperature than the sample, which cools via the exchange gas in the container. Thus at the appropriate temperature the nitrogen inside the sample can will condense onto the walls of the container, where it won't be observed in the scattering experiment when the incident and scattered beams are suitably restricted. On warming, on the other hand, the walls will be at a higher temperature than the sample. The nitrogen will then evaporate from the walls and condense onto the sample, where the polycrystalline $N_2$ can scatter neutrons into the detector.

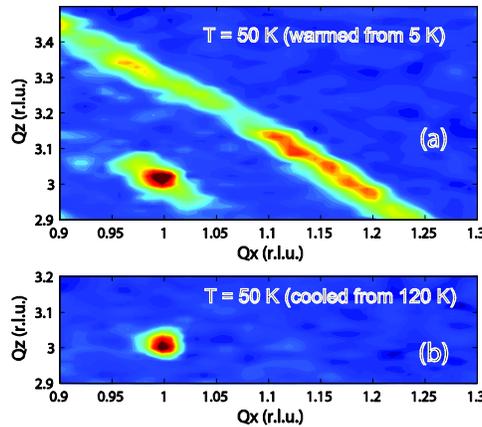

**Figure 1. Contour plot of the ring of scattering from β-$N_2$ for $|\mathbf{Q}|$=2.12 Å$^{-1}$ that appears on warming. The intensity and range of temperature where the ring is observed is dependent on the warming rate. (b) On cooling the ring is not observed.**

The scattering of the solid $N_2$ also shows strong time dependence. At 58 K, the intensity of the ring loses half its value in less than an hour. We note that the transition temperature $T_{\alpha \leftrightarrow \beta}$ strongly depends on the residual order above the transition [4], and it is also reported that the α↔β transition has hysteresis and takes as long as 10 hours for the system to equilibrate and reliably measure the intrinsic $T_{\alpha \leftrightarrow \beta}$ [5]. Hence for the present data the warming and cooling rates were too fast to produce an equilibrium phase. This explains the time dependence and why the onset temperature of the β-$N_2$ varies with different warming rates (figure 2(a)).
In the case of $N_2$ contamination, accidently scanning across the solid $N_2$ rings is very likely. The unexpected solid $N_2$ rings can have intensities comparable to magnetic Bragg peaks of a small sample, which is often used in elastic neutron scattering. The resultant confusion can be even greater if a Bragg point happens to fall in the middle between two $N_2$ powder rings which, regardless of the scan directions, would yield two symmetric peaks that can easily be mistaken as incommensurate magnetic scattering. We remark, though, that the structural origin of this scattering was confirmed by polarized neutron diffraction measurements on BT-7. Even a contour plot wouldn't necessarily reveal the rings because of their decay of intensities over time. A complete listing of $d$ spacings and calculated intensities of solid α-$N_2$ and β-$N_2$ can be found in reference [6].

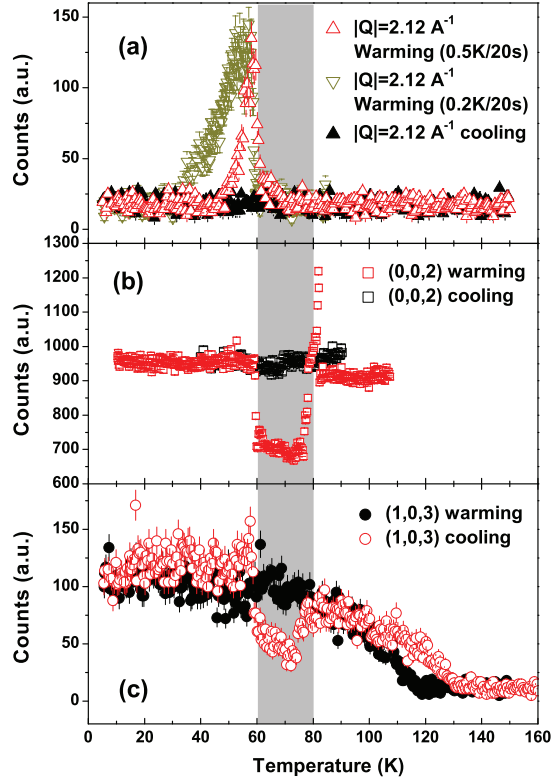

**Figure 2.** Temperature dependence of the peak intensity of (c) the Bragg peak of β-$N_2$ at $|Q|=2.12$ Å$^{-1}$, and (b) the (0,0,2) nuclear Bragg peak, (c) the (1,0,3) magnetic Bragg peak from the sample on warming (open red and green symbols) and on cooling (solid black symbols).

Another effect of $N_2$ contamination comes from the large thermal expansion of $N_2$ both at the α↔β transition as well as below ≈60 K just before it liquefies [7]. The thermal expansion may affect the alignment of the sample and thus the intensities of Bragg peaks if the sample is not rigidly mounted. An example of the intensity change of the scattering on warming and cooling is shown in figure 2 (b) and (c). These data were taken without any motor movement, so that the peak intensity vs. temperature is displayed. For temperatures above 80 K the peak intensity of the (002) structural peak (figure 1(b)) is smooth, and the development of the (103) magnetic peak occurred at about 130 K upon cooling (figure 1(c)). On cooling, typically the data are smooth and continuous all the way to low temperatures. However, on warming we observed sharp and dramatic changes in the intensities of both nuclear and magnetic Bragg peaks between 60 K and 80 K. We were able to observe some small changes in the Bragg intensities on very slow cooling, but the dramatic changes were only observed on warming for the reason discussed above. Comparable data were obtained on other nuclear Bragg peaks. As revealed by rocking scans of Bragg peaks (figure 3(a)), these alignment changes are ≈0.1-0.2°, which is enough of a misalignment of the crystal to perturb the Bragg peak intensities.

All the data with red (open) symbols in Figure 2 were taken at the same warming rate in the same CCR. The data clearly show that the sudden change in the single crystal Bragg intensity happens when the powder rings disappear. Above the boiling point of liquid $N_2$ the Bragg peaks recover their intensities. To establish with certainty that all these effects are due to $N_2$ in the sample can, we carried out measurements with pure He sealed in the can, and then with air. All the anomalous phenomena were observed with $N_2$ in the sample can, and disappeared when the He-sealed sample was measured, as shown in figure 3.

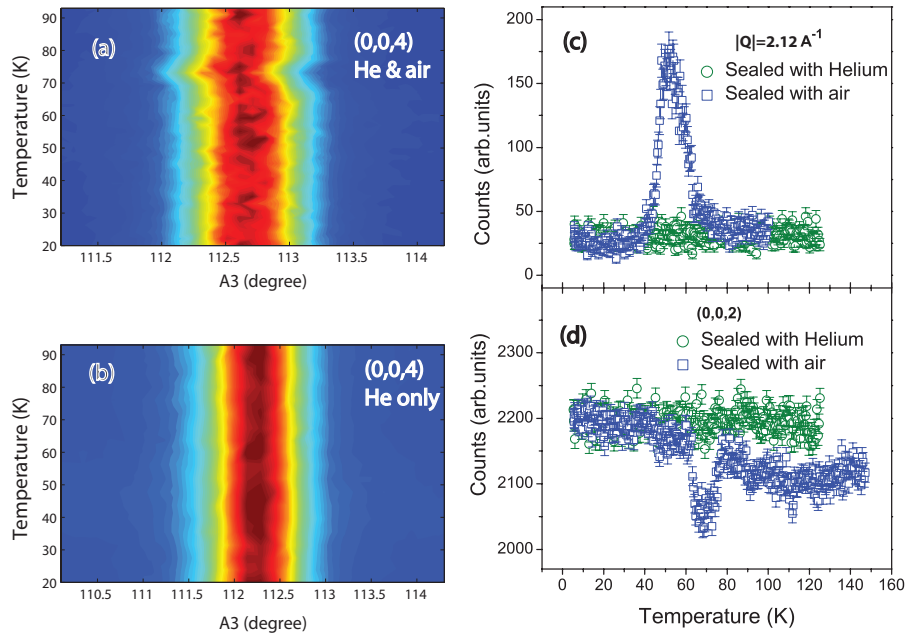

**Figure 3.** (a) Rocking scans (A3) of the (0,0,4) structural Bragg peak for the crystal, which was sealed with tiny amount of air inside. There is a sudden shift in position in the temperature range between 60 K and 80 K. (b) This anomalous shift in orientation of the crystal does not occur if the sample can is sealed with pure He. (c, d) Peak intensity for the ring and (0,0,2) Bragg peak on warming, with the sample can sealed in air (squares), or with pure helium gas (circles). The anomalous scattering is only observed when air is in the sample can.

## 4. Conclusion

In summary, during elastic neutron scattering measurements we have observed anomalous scattering due to $N_2$ condensing on the sample. The solid phases of $N_2$ can produce powder Bragg peaks of comparable intensity to the magnetic peaks, and also the thermal expansions during $N_2$ phase transitions can affect the alignment of the crystal. These problems can occur in a sealed sample container if for some reason the helium exchange gas is not sufficiently pure, or for top-loading sample environment systems that develop a leak. It is hoped that the present results will alert investigators to quickly identify these problems when they occur in the future.

**Acknowledgements** NPB is supported by the Center for Nanophysics and Advanced Materials.